# Discrete Transformation Elasticity: An Approach to Design Lattice-based Polar Metamaterials

Yangyang Chen, Hussein Nassar and Guoliang Huang[*]

Department of Mechanical and Aerospace Engineering, University of Missouri, Columbia, MO, 65211, USA

The transformation method is a powerful tool for providing the constitutive parameters of the transformed material in the new coordinates. In transformation elasticity, a general curvilinear change of coordinates transforms conventional Hooke's law into a different constitutive law in which the transformed material is not only anisotropic but also polar and chiral and no known elastic solid satisfies. However, this state-of-the-art description provides no insight as to what the underlying microstructure of this transformed material could be, the design of which is a major challenge in this field. The study aims to theoretically justify the fundamental need for the polar material by critically revisiting the discrete transformation method. The key idea is to let transformation gradient operate not only on the elastic properties but on the underlying architectures of the mechanical lattice. As an outstanding application, we leverage the proposed design paradigm to physically construct a polar lattice metamaterial for the observation of elastic carpet cloaking. Numerical simulations are then implemented to show excellent cloaking performance under different static and dynamic mechanical loads. The approach presented herein could promote and accelerate new designs of lattice topologies for transformation elasticity in particular and is able to be extended for realizing other emerging elastic properties and unlocking peculiar functions including statics and dynamics in general.

*Corresponding author: G Huang (Email: huangg@missouri.edu)

**Introduction**

The theory of composites, broadly defined, continues to inform design procedures of artificial materials [1]. An important thematic problem in that context is how to design materials with target elastic properties out of a limited number of elastic phases, with specific volume fractions, and using certain fabrication processes, e.g., mixing, layering, or 3D printing. The advent of metamaterials has presented the theory of man-made structured composites with new challenges where the target properties are unusual in the sense that they break certain conventional symmetries of the constitutive law, e.g., breaking Hooke's law [2-6]. The transformation method provides an efficient tool to find necessary material properties and distribution when a wave pattern is prescribed in a space. In transformation acoustics, for instance, the design of acoustic "invisibility" cloaks naturally calls for the use of "anisotropic fluids", i.e., acoustic media where stress is not necessarily hydrostatic [7-9]. Remarkably, anisotropic fluids can be 3D printed out of a single solid phase in a lattice form, in that case, anisotropic fluids are better known as "pentamode materials" [9-12]. In transformation elasticity, anisotropy is not enough, and a curvilinear change of coordinates transforms Hooke's law into different nonstandard constitutive laws [13]. Under the Milton–Briane–Willis gauge, the transformed material allows symmetric stresses and couples stresses related to displacements, known as Willis materials [13-15]. On the other hand, the Brun–Guenneau–Movchan gauge requires the transformed material with nonstandard elasticity tensors **c** that exhibit polar (i.e., breaking minor symmetry with $c_{ijkl} \neq c_{jikl}$) and, in some cases, chiral (i.e., lacking mirror symmetry) [13,16-21]. This observation begs the question of how to design elastic lattice metamaterials with targeted anisotropic, polar, and chiral elasticity tensors. In this paper, we answer that question for a wide range of materials in two space dimensions.

Lattice metamaterials, including pentamodes and other structured materials investigated hereafter, are a class of artificial cellular materials made out of interconnected beam elements organized at different scales to precisely tailor material properties. Evidently, lattice materials have applications that extend beyond the realm of cloaking, in particular, they provide designs for lightweight structures [22,23], bone replacements [24,25], energy absorbers [26-28], nanomaterials with ultrahigh strength, damage tolerance, and stiffness [29,30], and of other multifunctional materials [31-33]. Lattice metamaterials are, in a sense, universal for the realization of extreme and unusual properties. As a matter of fact, pentamodes have been originally introduced by Milton and Cherkaev as a way to prove that any elasticity tensor can be realized with an appropriate mixture of lattice materials [35]. In general, such mixtures are highly intricate and involve several interpenetrating lattices making them impractical. In some cases, however, a single lattice suffices to achieve the desired properties; such is the case of pentamodes in transformation acoustics. Here, we investigate how transformed elastic materials can be fashioned out of single-lattice metamaterials and propose a rational approach for their inverse design starting from the desired properties and resulting in 3D-printable architectures.

The desired elasticity tensor **c** of transformed elastic materials are due to two factors: the elasticity tensor **C** of the background medium before transformation; and, the gradient **F** of a space-warping transformation $\phi$ which maps a point in a virtual space to another point in a real space and therefore defines the geometry of the transformed material [13]. In a recent breakthrough, elastic cloaks have been designed by developing polar elasticity for two particular cases of transformations where (*i*) $\phi$ is conformal, that is **F** is shear-deformation-free and proportional to a rotation [18,20]; and (*ii*) $\phi$ is radially symmetric and, in particular, **F** is rotation-free [17,19,21]. By contrast, the present design paradigm holds for arbitrary transformation gradients **F** combining nonuniform stretch, shear, and rotation to realize anisotropic polar materials with chiral properties. The key idea is to let **F** operate, not only on the elastic properties to change them from **c** to **C**, but on the underlying architectures themselves. In other words, instead of looking for architectures of the transformed materials with the unusual tensor **c**, we first target the architectures of the background medium with conventional tensor **C**. The found architectures are then transformed by **F** into a lattice material which automatically exhibits the sought-after tensor **c**. The transformation rules according to which **F** operates on a lattice material to generate another lattice material constitute what we refer to as "Discrete Transformation Elasticity". A theoretical treatment of discrete transformation elasticity was first proposed by Guevara Vasquez et al. but did not lead to feasible designs [36]; another relevant contribution, is the one by Bückmann et al. [37], where direct lattice transformations have been suggested but ignored the polar character of the required tensors. Here, we complete these efforts under the framework of polar elasticity and propose feasible designs based on rigorous derivations taking into account the full tensorial character of the equations of elasticity that can be anisotropic, polar, and chiral simultaneously. As an outstanding application, we leverage the proposed design paradigm to physically construct a polar lattice metamaterial for the observation of elastic carpet cloaking. Numerical simulations are then conducted to show excellent cloaking performance under shear and pressure, static and dynamic loads.

**Continuum and discrete transformation elasticity**

Consider a general mapping $\mathbf{x} = \phi(\mathbf{X})$ that transforms a virtual (original) domain $\{\mathbf{X}\}$ into a physical (transformed) domain $\{\mathbf{x}\}$ (see Fig. 1). As for the range of possible transformations, $\phi$ encompasses any combinations of stretch, shear, and rotation with the transformation gradient **F** decomposed by $\mathbf{F} = \mathbf{VR}$ where **R** is orthogonal (rotation tensor) and **V** is symmetric positive definite (deformation tensor), as shown in Fig. 1. The virtual and physical domains are occupied by elastic continuum media which, under an external load, are displaced by fields $\mathbf{U}(\mathbf{X})$ and $\mathbf{u}(\mathbf{x})$, respectively. We are interested in determining the constitutive properties of $\{\mathbf{x}\}$, or even its microstructure, that let the displacement field be warped according to the same transformation, that is, such that $\mathbf{u}(\mathbf{x}) = \mathbf{U}(\mathbf{X})$. To do so, it is particularly insightful to interpret the fields $\mathbf{u}(\mathbf{x})$ and $\mathbf{U}(\mathbf{X})$ as two different, but equivalent, sets of generalized Lagrangian coordinates, namely such that

$$\int_{\{\mathbf{X}\}} L(\nabla \mathbf{U}) \mathrm{d}\mathbf{X} = \int_{\{\mathbf{x}\}} \ell(\nabla \mathbf{u}) \, \mathrm{d}\mathbf{x}, \qquad (1)$$

where $L$ and $\ell$ are the strain energy densities over $\{\mathbf{X}\}$ and $\{\mathbf{x}\}$, respectively. The change of coordinates formula, together with the chain rule, then yield $\ell(\nabla \mathbf{u}) = L(\nabla \mathbf{u} \mathbf{F})/J$ with $\mathbf{F} = \mathrm{d}\mathbf{x}/\mathrm{d}\mathbf{X}$ and $J = \det \mathbf{F}$ being the transformation gradient and its determinant. Accordingly, it is possible to warp the displacement field as long as $\{\mathbf{x}\}$ is composed of materials with the prescribed strain energy $\ell$. In terms of the constitutive properties, when the original domain $\{\mathbf{X}\}$ has an elasticity tensor $\mathbf{C}$, the transformed domain $\{\mathbf{x}\}$ has an elasticity tensor $\mathbf{c}$ with $c_{ijkl} = F_{jm} F_{ln} C_{imkn}/J$. A close inspection of the foregoing relation shows that the transformed elasticity tensor $\mathbf{c}$ is unconventional in at least three regards: (*i*) it is polar in the sense that it lacks the minor symmetry; (*ii*) it is degenerate in the sense that it admits a number of zero modes; (*iii*) it is chiral in the sense that it lacks mirror symmetry (in 2D) when $\mathbf{F}$ is not rotation-free. Materials with such unusual properties are unavailable and lattice-based designs, in few particular cases, have only recently been found [17-21]. This state of affairs has significantly impeded progress in transformation elasticity in comparison to its optics or acoustics counterpart.

Here, we solve the material design problem by fully embracing a discrete lattice-based transformation (see Fig. 2). Thus, we discretize $L$ and $\ell$ as if they represented the strain energy densities of two lattices. Without loss of generality, we start the architecture of the background medium with a triangle virtual lattice in a periodic configuration. The lattice is made of a set of massless springs connecting to the mass nodes (Fig. 2). Hence, we assume

$$L(\nabla \mathbf{U}) = \sum_{m,n,p} K^p \langle \nabla \mathbf{U}(\mathbf{X}^p_{m,n} - \mathbf{X}_{m,n}), \mathbf{S}^p \rangle^2 / A, \qquad (2)$$

the strain energy per unit cell area $A$ of a set of springs (index $p$) of constants $K_p$ and of direction $\mathbf{S}^p = \frac{\mathbf{X}^p_{m,n} - \mathbf{X}_{m,n}}{\|\mathbf{X}^p_{m,n} - \mathbf{X}_{m,n}\|}$ connecting node $\mathbf{X}_{m,n}$ of a given unit cell (index $m, n$) to its neighboring nodes $\mathbf{X}^p_{m,n}$. Then, per the above identity relating $L$ and $\ell$, domain $\{\mathbf{x}\}$ will have the same behavior as a lattice whose strain energy is

$$\ell(\nabla \mathbf{u}) = \sum_{m,n,p} k^p \langle \nabla \mathbf{u}(\mathbf{x}^p_{m,n} - \mathbf{x}_{m,n}), \mathbf{s}^p \rangle^2 / a, \qquad (3)$$

with $\mathbf{x}_{m,n} = \mathbf{F} \mathbf{X}_{m,n}$, $a = AJ$, $k^p = K^p$ and $\mathbf{s}^p = \mathbf{S}^p$. In other words, the lattice of $\{\mathbf{x}\}$ can be easily deduced from that of $\{\mathbf{X}\}$ by (*i*) applying the transformation gradient $\mathbf{F}$ to the nodes of the lattice (Fig. 2) while (*ii*) leaving the springs constants and directions, as they were (Fig. 2). In particular, note that the lattice vectors anchored at the nodes, such as $\mathbf{X}^p_{m,n} - \mathbf{X}_{m,n}$, and the spring orientation $\mathbf{S}^p$ do not transform in the same fashion; indeed, $\mathbf{s}^p \neq \frac{\mathbf{x}^p_{m,n} - \mathbf{x}_{m,n}}{\|\mathbf{x}^p_{m,n} - \mathbf{x}_{m,n}\|}$. Last, to maintain physical contact between the mass nodes and the misaligned springs, the mass nodes must assume a finite size and become rigid bodies (see the transformed polar lattice in Fig. 2). This has the unintended effect of introducing an extra rotational degree of freedom which must be suppressed using an external ground or other rotational resonant structures. Together, the misalignment and ground or rotational resonant mechanisms, are responsible for the unconventional properties of the transformed tensor **c**.

It is important here to stress that the original domain {**X**} need not be a lattice, and only to behave like one. In particular, all isotropic materials with a Poisson's coefficient equal to 1/3 behave like a triangular truss and can therefore be transformed in this fashion. More generally, square, rectangular, and oblique lattices with anisotropic effective tensors **C** can be transformed in the same manner. Conversely, the availability of a lattice representation of the original domain is the only limitation weighing on the present approach. Note also that this study only focuses on one form of $\ell(\nabla \mathbf{u})$ for the lattice in {**x**} where rotations of rigid masses are suppressed, while other forms of $\ell(\nabla \mathbf{u})$ could also be feasible, e.g. enabling rotations of rigid masses by connecting them with grounded rotational springs [17].

**A polar lattice for elastic carpet cloaking**

Having introduced the continuum and discrete transformation elasticity, it is time to illustrate the usefulness of the suggested paradigm to the design of polar lattices displaying nonstandard elastic properties never realized before. Our particular interest is to construct an elastic carpet cloak with material properties not changing in space, beneficial to manufacturing and practical applications. To facilitate this design criterion, we apply a geometric transformation that linearly compresses a triangular virtual space (shaded green area) in a way to a physical space surrounded by a polygon (shaded yellow area), where boundaries of the background medium left unchanged (see Fig. 3a). The resulted triangular void along the free boundary of a two-dimensional elastic medium can then be concealed by the transformed material with desired mass density and elastic tensor **c**. In particular, the transformation $\phi$ between the virtual {**X**} and physical {**x**} coordinates reads

$$x = X, \quad y = \frac{\tan(\alpha) - \tan(\theta)}{\tan(\alpha)} Y + \left[l_a - \mathrm{sgn}(x) x\right] \tan(\theta), \tag{4}$$

where $\alpha$ and $\theta$ denotes angles of the triangle in the virtual space and the triangular void in the physical space, and $l_a$ is the length of the bottom leg of the triangle in the virtual space. We then obtain the deformation gradient of $\phi$ as

$$\mathbf{F} = \begin{bmatrix} 1 & 0 \\ -\mathrm{sgn}(x)\tan(\theta) & 1 - \dfrac{\tan(\theta)}{\tan(\alpha)} \end{bmatrix}. \tag{5}$$

Based on polar decomposition, we have

$$\mathbf{V} = \begin{bmatrix} p_1 \cos^2(\varphi) + p_2 \sin^2(\varphi) & (p_2 - p_1)\cos(\varphi)\sin(\varphi) \\ (p_2 - p_1)\cos(\varphi)\sin(\varphi) & p_2 \cos^2(\varphi) - p_1 \sin^2(\varphi) \end{bmatrix}, \quad \mathbf{R} = \mathbf{V}^{-1}\mathbf{F}, \tag{6}$$

where $p_{1,2} = \dfrac{1 + F_{21}^2 + F_{22}^2}{2} \mp \sqrt{\left(\dfrac{1 - F_{21}^2 - F_{22}^2}{2}\right)^2 + F_{21}^2}$ and $\tan(-2\varphi) = \dfrac{2F_{21}}{1 - F_{21}^2 - F_{22}^2}$. It can be clearly seen that the stretch is nonuniform, as $V_{11} \neq V_{22}$. The physical coordinate is eventually compressed along the y-direction and left unchanged along the x-direction. In addition, the transformation comprises a shear deformation, since $V_{12} = V_{21} \neq 0$. Finally, the deformed grid needs to experience a rigid rotation $\mathbf{R} \neq \mathbf{I}$ to realize the transformation gradient $\mathbf{F}$.

Assume the material in the virtual space is isotropic with Lamé parameters $(\lambda, \mu)$ and mass density $\rho_0$. Applying continuum transformation elastodynamics [13], the mass density and constitutive relations of the transformed material that conceals the triangular void (shaded yellow area) read

$$\rho = \rho_0 \mathbf{A}, \tag{7}$$

$$\begin{bmatrix} \sigma_{11} \\ \sigma_{22} \\ \sigma_{12} \\ \sigma_{21} \end{bmatrix} = \begin{bmatrix} (\lambda+2\mu)A & \lambda & (\lambda+2\mu)C & 0 \\ \lambda & (\lambda+2\mu)/A+\mu BC & (\lambda+\mu)B & \mu c \\ (\lambda+2\mu)C & (\lambda+\mu)B & \mu/A+(\lambda+2\mu)BC & \mu \\ 0 & \mu C & \mu & \mu A \end{bmatrix} \begin{bmatrix} e_{11} \\ e_{22} \\ e_{12} \\ e_{21} \end{bmatrix}, \qquad (8)$$

where $A = \dfrac{\tan(\alpha)}{\tan(\alpha)-\tan(\theta)}$, $B = -\mathrm{sgn}(x)\tan(\theta)$ and $C = AB$. The matrix in the right-hand side of Eq. (8) represents the transformed elastic tensor **c**. Eq. (7) indicates that the total mass of an area in the virtual space is equal to the total mass of its transformed area in the physical space. On the other hand, according to Eq. (8), the transformed elastic material for carpet cloaking should be anisotropic, as $c_{1111} \neq c_{2222}$ and $c_{1212} \neq c_{2121}$ due to the nonuniform stretch and shear in the transformation. Furthermore, the transformed material also needs to be chiral ($c_{1211} \neq c_{2111}$ and $c_{1222} \neq c_{2122}$), and the material is not centrosymmetric since the transformation comprises a rotation after a general deformation (nonuniform stretch and shear). If we are able to design a real medium equivalent to the transformed material, called "cloaking material", described in Eqs. (7) and (8), then we are able to accomplish the cloaking behavior for elastic waves in the physical space. However, efforts so far provide no insight as to what the underlying microstructure of this cloaking material could be, the design of which is a major challenge in this field.

We tackle the inverse problem of designing the cloaking material using discrete transformation elasticity, from which a mass-spring lattice is suggested for the non-standard constitutive properties in Eqs. (7) and (8). To do this, the isotropic background medium with the Poisson's ratio assumed as 1/3 ($\lambda = \mu$) is first represented by a triangle virtual lattice with massless springs connecting to mass nodes (see the left figure in Fig. 3b). The length of the springs is denoted by $a$, and the spring constants ($k_1$, $k_2$, $k_3$) and mass $m$ of the triangular lattice can be determined from the isotropic material properties by $k_1 = k_2 = k_3 = \dfrac{4\mu}{\sqrt{3}}$ and $m = \dfrac{\sqrt{3}\rho_0 a^2}{2}$. Performing the proposed discrete transformation, mass nodes of the virtual lattice are mapped to new locations in the transformed space following the transformation $\phi$ (see Points 0 – 6 in Fig. 3b before and after the transformation), and directions of the connecting springs remain as they were in the virtual lattice. To make masses and springs physically connected, we manually adjust spring lengths and shapes and sizes of the rigid masses. Note that the constructed lattice configuration can be arbitrary as long as the configuration is periodic and free of overlap connections between masses and springs. In the design, we select rigid line masses, and enforce springs $k_1$ and $k_2$ to the locations that transformed from $\phi$ and connected to one end of the rigid mass on Point 0. By doing this, rigid masses must be in alignment with the springs $k_1$ for periodicity, and their lengths are determined by connecting springs $k_1$ and $k_2$ to the other ends of the rigid masses on Points 1 and 2 (see the right figure in Fig. 3b). Finally, the spring $k_3$ is adjusted into the transformed lattice by connecting Point 0 to one end of the rigid mass on Point 6. In the design, all contacts are assumed to be hinge-like, and rotations of masses are suppressed.

Now we focus on the homogenized elastic response of the transformed lattice to validate the design. Using the volume average approach [17], the effective elastic constants of the transformed spring-mass system can be analytically obtained in terms of $\theta$, $k_1$, $k_2$, and $k_3$ for $x > 0$ as (See Supplementary Information for details)

$$\begin{aligned}
&\bar{c}_{1111} = \bar{A}(16k_1 + k_2 + k_3),\ \bar{c}_{1122} = \bar{A}(3k_2 + \bar{B}k_3),\ \bar{c}_{1112} = \bar{A}(-16\tan(\theta)k_1 - \sqrt{3}k_2 + \bar{C}k_3),\\
&\bar{c}_{1121} = \sqrt{3}\bar{A}(-k_2 + k_3),\ \bar{c}_{2222} = \bar{A}(9k_2 + 3\bar{C}^2 k_3),\ \bar{c}_{2212} = \sqrt{3}\bar{A}(-3k_2 + \bar{C}^2 k_3),\\
&\bar{c}_{2221} = 3\bar{A}(-\sqrt{3}k_2 + \bar{C}k_3),\ \bar{c}_{1212} = \bar{A}(16\tan^2(\theta)k_1 + 3k_2 + \bar{C}^2 k_3),\ \bar{c}_{1221} = \bar{A}(3k_2 + \bar{B}k_3),\\
&\bar{c}_{2121} = 3\bar{A}(k_2 + k_3),
\end{aligned} \qquad (9)$$

where $\bar{A} = \dfrac{1}{8\left[\sqrt{3} - \tan(\theta)\right]}$, $\bar{B} = 3 - 2\sqrt{3}\tan(\theta)$, and $\bar{C} = \sqrt{3} - 2\tan(\theta)$. Applying the conditions enforced by the virtual lattice: $k_1 = k_2 = k_3 = \dfrac{4\mu}{\sqrt{3}}$ and $\lambda = \mu$, effective elastic constants of the mass-spring lattice in Eq. (9) can be exactly the same with the transformed material parameters in Eq. (9), coincide with the predictions in discrete transformation elasticity.

It should be worth mentioning that the lattice we constructed through discrete transformation elasticity naturally admits a zero-mode $\mathbf{e}_0 = \mathbf{E}_0 \mathbf{F}^{-1}$, where $\mathbf{E}_0 = \begin{bmatrix} 0 & 1 \\ -1 & 0 \end{bmatrix}$. Physically, the rigid rotation in the virtual lattice is transformed into a zero-mode in the transformed lattice. As shown in Fig. 3c, the motions on Points 1-6 are identical before (the motion is a rigid rotation) and after (the motion is a zero-mode) the transformation, exactly satisfying the displacement gauge $\mathbf{U}(\mathbf{X}) = \mathbf{u}(\mathbf{x})$ applied in the transformation.

**Microstructure realization of the polar metamaterial**

Informed by the transformed mass-spring lattice, we now explore ways to numerically design a polar metamaterial to confirm the extent to which a candidate metamaterial microstructure meets the requirements of a cloaking material and is ready to be fabricated. Figures 4a and 4b illustrate the metamaterial design, where a hard material (blue area) is employed for constructing rigid masses, and a soft material (yellow area) functions as springs. The geometry of the hard material can be flexible to meet the mass density requirement. However, it should at least meet the requirement in such a way that the hard material should connect to the soft material at the three points located on the red dotted line (See detailed geometric parameters of this design in Supplementary Information). To suppress the rotational motion of each mass, we introduce a rod connecting the mass to the ground (see Fig. 4c). The rod near the ground is indented to significantly reduce its bending stiffness along $u_1$ and $u_2$ directions. As a result, the rotational stiffness of the rod is much greater than its bending stiffness. Therefore, rotation of the supported mass is efficiently suppressed and only in-plane translation is allowed. The conventional diamond-shaped bar made of the soft material is selected to mimic the spring (Fig. 4b), which leads to negligible bending moments at its ends when the bar twists around the mass. By varying diameters $d_1$, $d_2$, and $d_3$, we can realize different effective spring constants $k_{eff} = k_1 = k_2 = k_3$ for the transformed lattice.

Having selected geometric and material parameters of the metamaterial, it is our interest to determine the effective elastic constants for desired cloaking application. We numerically calculate the effective elastic constants based on the volume average approach similar to that employed analytically [17]. In the approach, the mechanical response of the metamaterial and the effective medium is considered equivalent if the strain energy density of the former is equal to that of the latter. To obtain the effective **c**, ten independent numerical tests are performed for solving the ten independent components in **c** dictated by the major symmetry ($c_{ijkl} = c_{klij}$). In each of the tests, we apply the rotational constraint on the left bottom boundary of the mass to mimic the required grounded connection. We prescribe displacements on the boundaries of the unit cell to induce different strain states. Other boundaries are left free. During the simulations, we first individually apply the four uniform strain states: two uniaxial and two shear strains ($e_{11}$, $e_{22}$, $e_{12}$, or $e_{21}$), and then apply the six mixed strains: ($\{e_{11}, e_{22}\}$, $\{e_{11}, e_{12}\}$, $\{e_{11}, e_{21}\}$, $\{e_{22}, e_{12}\}$, $\{e_{22}, e_{21}\}$, or $\{e_{12}, e_{21}\}$). The strain energy per unit cell is numerically calculated for each of the ten cases with the plane stress hypothesis, from which the ten independent elastic constants can be retrieved. As shown in Fig. 4d, effective elastic constants of the lattice metamaterial (orange bars) agree very well with the transformed parameters in Eq. (8) (blue bars), which demonstrates the validity of the microstructure design.

**Cloaking simulations**

With the proposed microstructure of the lattice-based metamaterial, we are ready to construct a carpet cloak for elastic wave propagation tests (see Fig. 4e). Since the lattice-based metamaterial is originated from the discrete transformation elasticity, the metamaterial can automatically match the transformed physical space without any geometric alterations. This procedure, therefore, provides significant advantages in building lattice topologies for transformation elasticity. The number of unit cells in the cloak is chosen sufficiently large so as to enable satisfactory cloaking performance but not too large so as to avoid impractical simulation times. In the design, we discrete 19 unit cells in the bottom row of the lattice, and tessellate them row-by-row with a total of 19 rows to create the carpet cloak. Note that the lattice metamaterial has seamless connections with the background material as shown in Fig. 4e. From a fabrication point of view, the total mass of the hard material can be tuned or adjusted by geometry modifications. In particular, special treatments are made for masses on the middle line of the cloak, where discontinuities emerge. We merge the two half masses into one mass and properly enlarge its area equal to the area of other masses.

In numerical simulations, Navier's equations in the background medium and lattice metamaterial are solved using COMSOL Multiphysics. Perfectly matched layers surrounding the background medium are adopted to suppress reflected waves from boundaries. Incident waves with Gaussian profiles in the direction of -60° are emitted to the cloak region (see Supplementary Information for other incident angles). Simulation results are shown in Fig. 5 for an incident pressure wave (Figs. 5a and 5b) and incident shear wave (Figs. 5c and 5d) at 400 Hz. Figures in the first column show the results when the lattice metamaterial is embedded in the background medium for cloaking. For reference, we perform the same simulations where the background medium is intact (the second column) and the void is non-coated (the third column). The divergence and curl of the displacement field are shown in Figs. 5a and 5c and Figs. 5b and 5d, respectively. It is clearly seen that the designed lattice metamaterial demonstrates excellent cloaking performance as it almost perfectly suppresses pressure, shear, and Rayleigh scattering due to the presence of the void. Specifically, the lattice metamaterial makes significate improvements to scattered shear waves from the void, which are more sensitive to defects compared with pressure waves, due to the shorter wavelength. Simulations are also conducted to study the cloaking performance of the lattice metamaterial under different incidence angles (See Supplementary Information). The results still demonstrate an excellent cloaking performance.

The lattice metamaterial we designed comprises no resonance structures, implying that the design could be operated at much broader frequency regions wherever the metamaterial is well within the continuum limit. To demonstrate this, we perform numerical simulations at different frequencies. As shown in Fig. 6, pressure waves are excited in the direction of -30°, and we compare the cloaking performance of the metamaterial (Figs. 6a and 6b: Divergence and curl of the displacement field) with the wave fields where the background medium is intact (Figs. 6c and 6d: Divergence and curl of the displacement field). In the figure, the first to fourth columns show the simulation results at 200, 400, 600, and 800 Hz, respectively. It can be clearly seen that good cloaking performances still retain at frequencies below 600 Hz, whereas, at 800 Hz, the shear wave field produced by the metamaterial cloak slightly deviates from that reflected from a flat boundary. To improve the performance, one needs to reduce the size of the metamaterial unit cell.

Besides the dynamic pressure and shear loads, the lattice metamaterial cloak can also conceal the void with static loads. In the static numerical tests, we fix the right boundary of the background plate and prescribe displacements on the left boundary of the background plate. Other boundaries are left free. The prescribed displacement is along either horizontal (Figs. 7a and 7b) or vertical (Figs. 7c and 7d) directions to induce elongation or shear in the background plate. Similar to Fig. 5, the first column in Fig. 7 shows the results when the lattice metamaterial cloak is embedded, and the second and third columns are for the intact background medium and the non-coated void. Again, the divergence and curl of the displacement field are shown in Figs. 7a and 7c and Figs. 7b and 7d, respectively. We can easily find the lattice metamaterial cloak significantly improves the pressure and shear fields when the background plate undergoes elongation. In addition, the triangular void is insensitive to the shear load, where the figures in the first to third columns are nearly identical. Finally, it is also worth mentioning that the modulus of the hard material can be reduced in real designs, whereas quantitative studies should be performed to ensure the cloaking performance.

**Conclusions**

In conclusion, we report a simple approach based on discrete transformation elasticity to design lattice-based polar metamaterials that can automatically satisfy constitutive requirements from arbitrary coordinate transformations. In particular, we engineer a lattice metamaterial that exhibits desired polarity, chirality, and anisotropy of a transformed continuum. Based on this polar metamaterial, an exact elastic carpet cloak is constructed and demonstrated numerically for concealing a triangular void. The research establishes a theoretical framework for tackling the inverse design problems of lattice-based materials targeting transformed macroscopic constitutive tensors, not only for those required by elastic cloaks. Given the fact that transformation elasticity is a highly flexible method for exploiting new material parameters by varying coordinate transformations or materials in the virtual space, the lattice-based metamaterials designed based on discrete transformation elasticity could foster a large range of control functions in dynamics and statics in general i.e. waveguiding, illusion, and shape morphing. Furthermore, the metamaterials designed based on this approach is ready to be used without geometric alterations or discontinuities. This approach, therefore, holds advantages over other methods in building lattice-based topologies related to transformation elasticity.

**Acknowledgements**

This work is supported by the Army Research Office under Grant No. W911NF-18-1-0031 with Program Manager Dr. Daniel Cole.

**Author contributions**

Y.C. and G.H. proposed the design concept; Y.C. and H.N. performed theoretical and numerical investigations; G.H. supervised the research; All the authors discussed the results and wrote the manuscript.

**Figures**

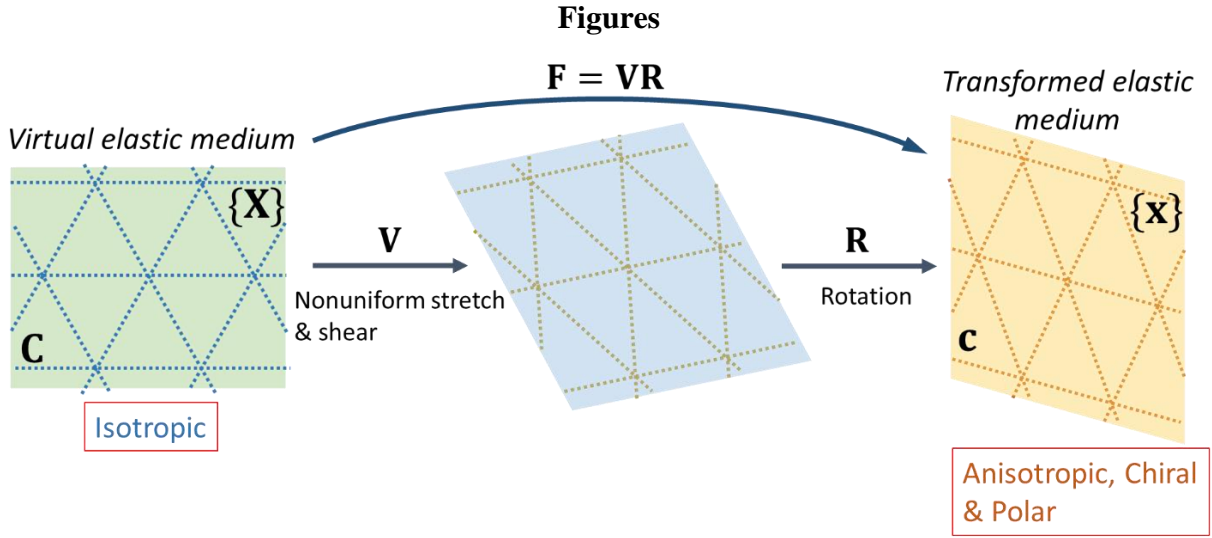

Figure 1. Schematic representation of continuum transformation elasticity. A continuum elastic medium occupies the virtual space {**X**} with the elastic tensor **C**. A transformed elastic medium occupies the physical space {**x**} with the elastic tensor **c**. The transformation gradient **F** comprises nonuniform stretch, shear, and rotation. When the virtual elastic medium is isotropic, the resulted transform elastic medium is anisotropic, chiral, and polar.

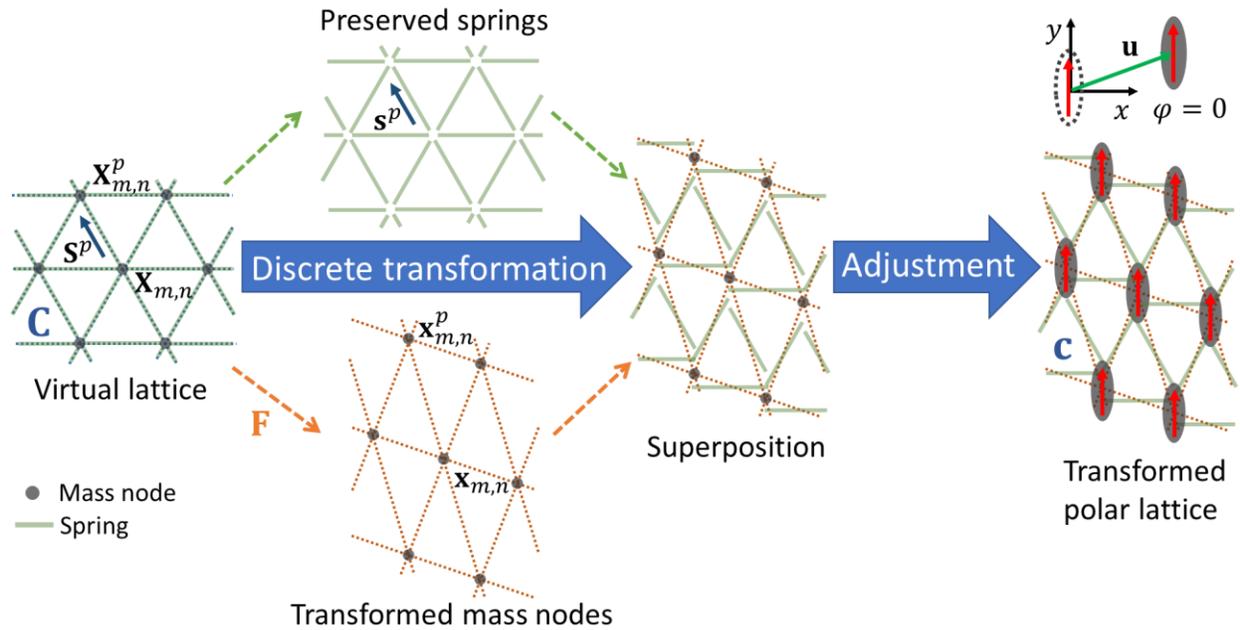

Figure 2. Schematic representation of discrete transformation elasticity. A virtual lattice with springs connected to mass nodes displays an effective modulus **C**. During the discrete transformation, mass nodes are transformed to new locations following **F**, while springs maintains their original directions. The transformed lattice is initially misaligned and does not have contact connections. The final transformed polar lattice is constructed by adjusting masses and springs to ensure contact and suppressing rotations of the masses.

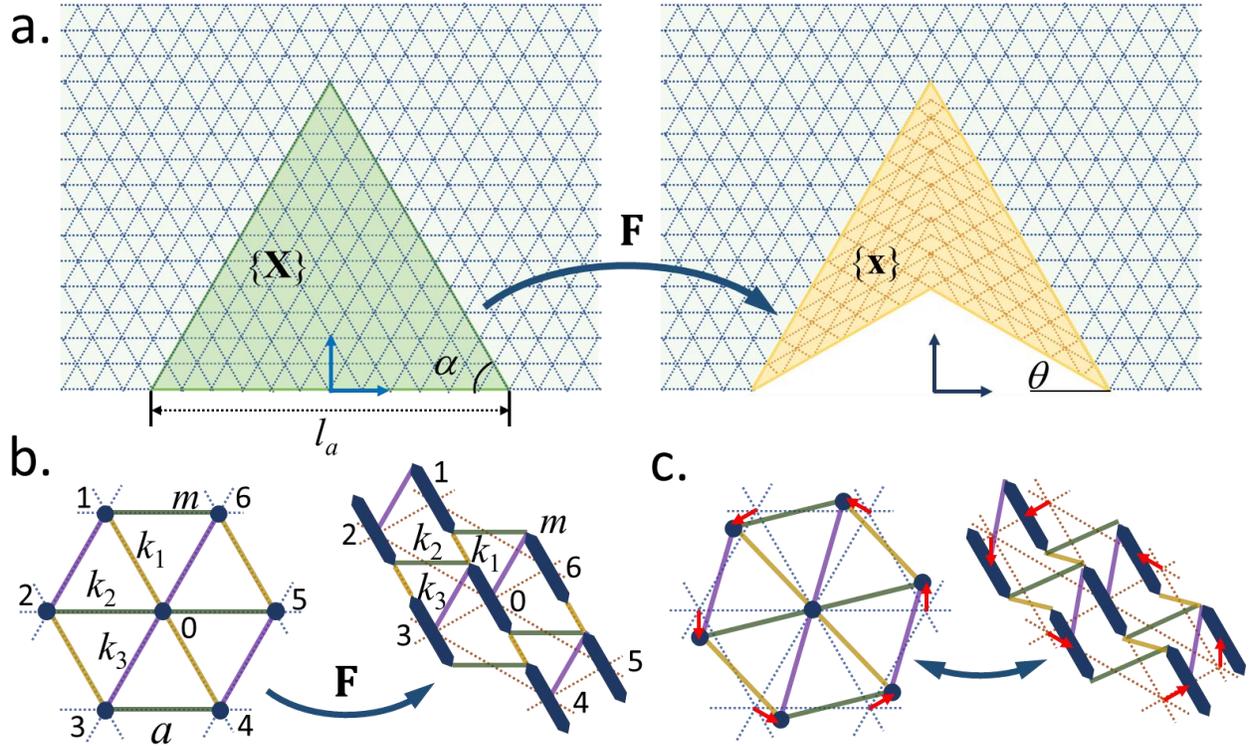

Figure 3. Design of the discrete mass-spring polar lattice for a carpet cloak. (a) Geometric transformation for realizing a carpet cloak concealing a triangular void; (b) The mass-spring lattice is designed using the discrete transformation. For the design, $\alpha = 60°$, and the lengths of springs, $k_1$, $k_2$, $k_3$, and the rigid masses are $a\left[1-\frac{2\sqrt{3}}{3}\tan(\theta)\right]$, $a\left[1-\frac{\sqrt{3}}{3}\tan(\theta)\right]$, $a\left[1-\frac{\sqrt{3}}{3}\tan(\theta)\right]$, and $\frac{2\sqrt{3}a}{3}\tan(\theta)$, respectively; (c) Zero-mode of the transformed lattice is equivalent to the rigid rotation of the virtual lattice.

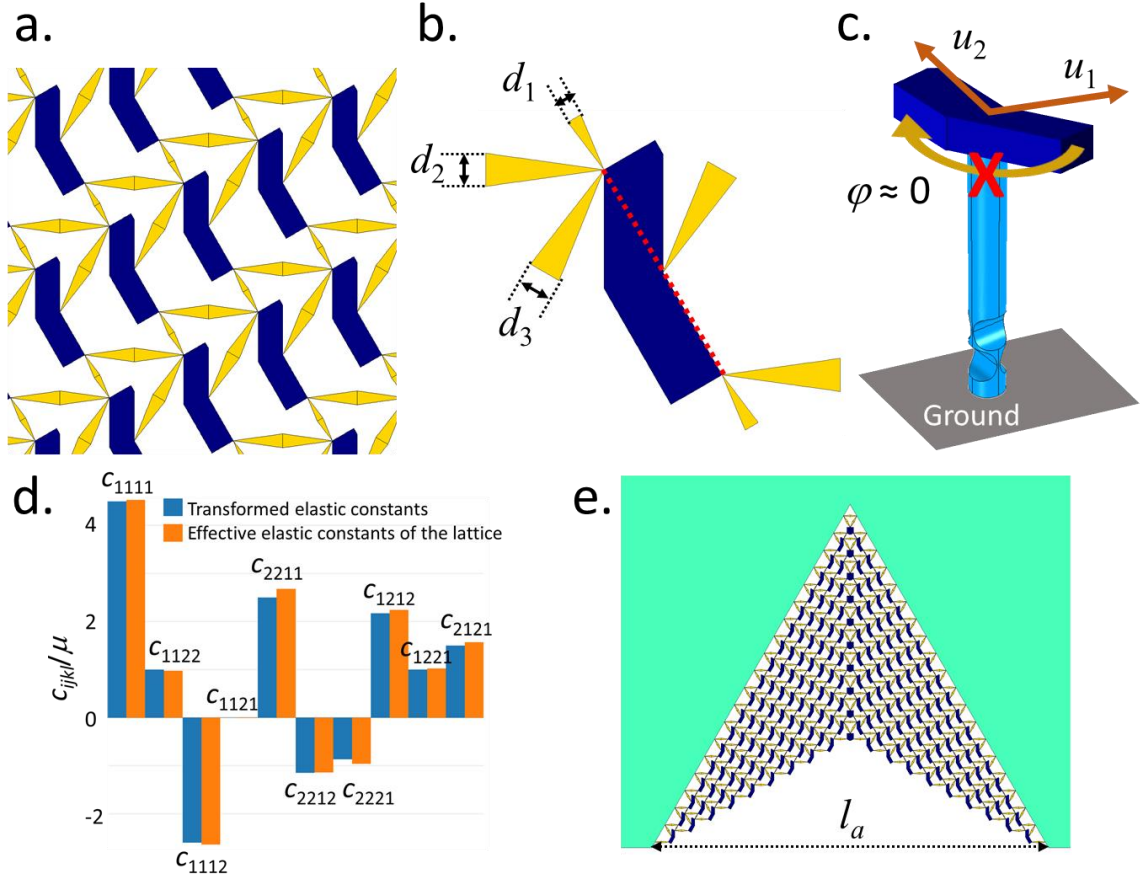

Figure 4. Microstructure realization of the lattice-based polar metamaterial. (a) The topology of the lattice metamaterial; (b) The unit cell of the lattice metamaterial. In the design, $a = 18$ mm, $\theta = 30°$, $d_1 = 0.38$ mm, and $d_2 = d_3 = 0.84$ mm. The Young's modulus and Poison's ratio of the soft and hard materials are 1 GPa, 0.33, 100 GPa, and 0.33, respectively; (c) The hard material is connected to the ground with a rod that supports near-zero rotation but nearly free in-plane translation; (d) Effective elastic constants of the lattice metamaterial in comparison with the desired transformed elastic constants; (e) A carpet cloak is constructed using the lattice metamaterial, where $l_a = 360$ mm. For the carpet cloaking, material parameters of the background material are selected as $\lambda = \mu = 9.14$ MPa and $\rho_0 = 1000$ kg/m$^3$. To satisfy the mass requirement, the mass density of the hard material is assumed as $\rho_h = 7035$ kg/m$^3$.

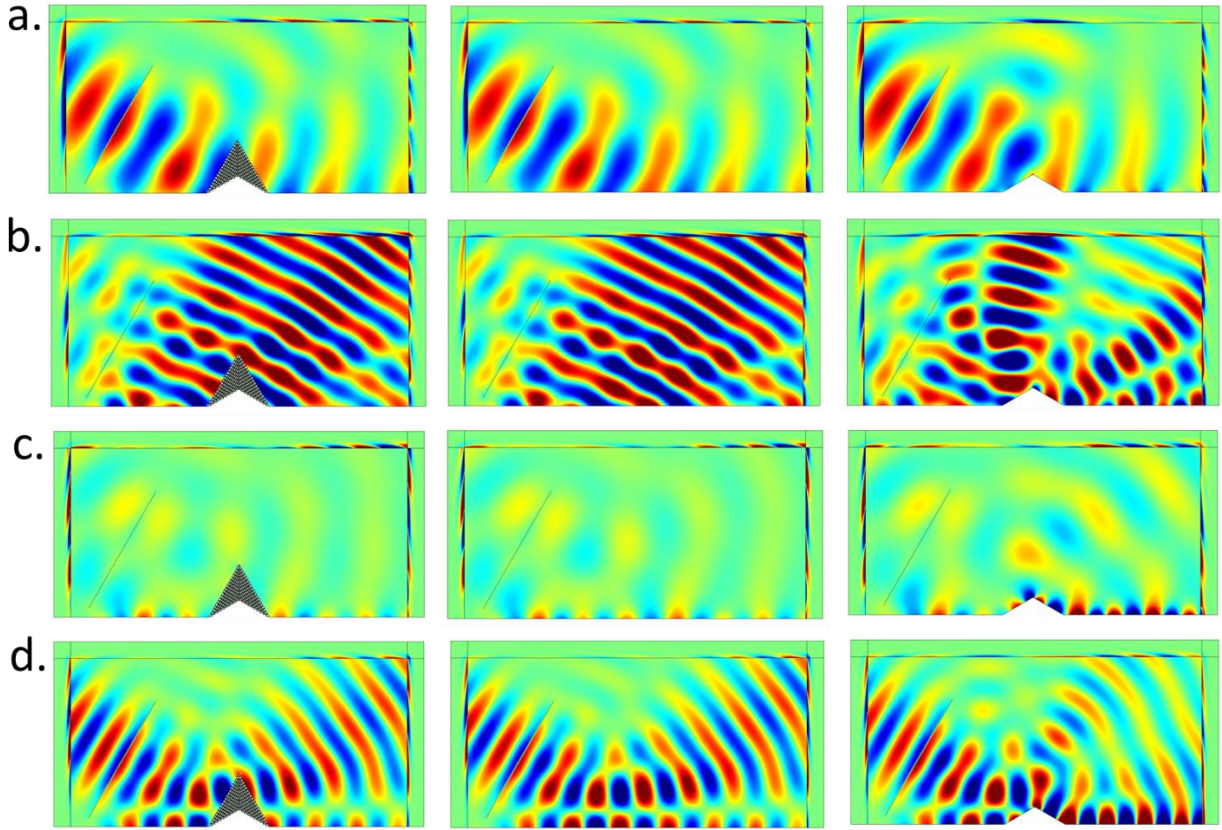

Figure 5. Numerical simulations of the carpet cloak based on the lattice-based polar metamaterial. Figures in the first column show the results when the continuum lattice is embedded in the background medium for cloaking. Figures in the second column show the results when the background medium is intact. Figures in the third column show the results when the void is non-coated. (a) Divergence of the displacement field with a pressure incidence; (b) Curl of the displacement field with a pressure incidence; (c) Divergence of the displacement field with a shear incidence; (d) Curl of the displacement field with a shear incidence.

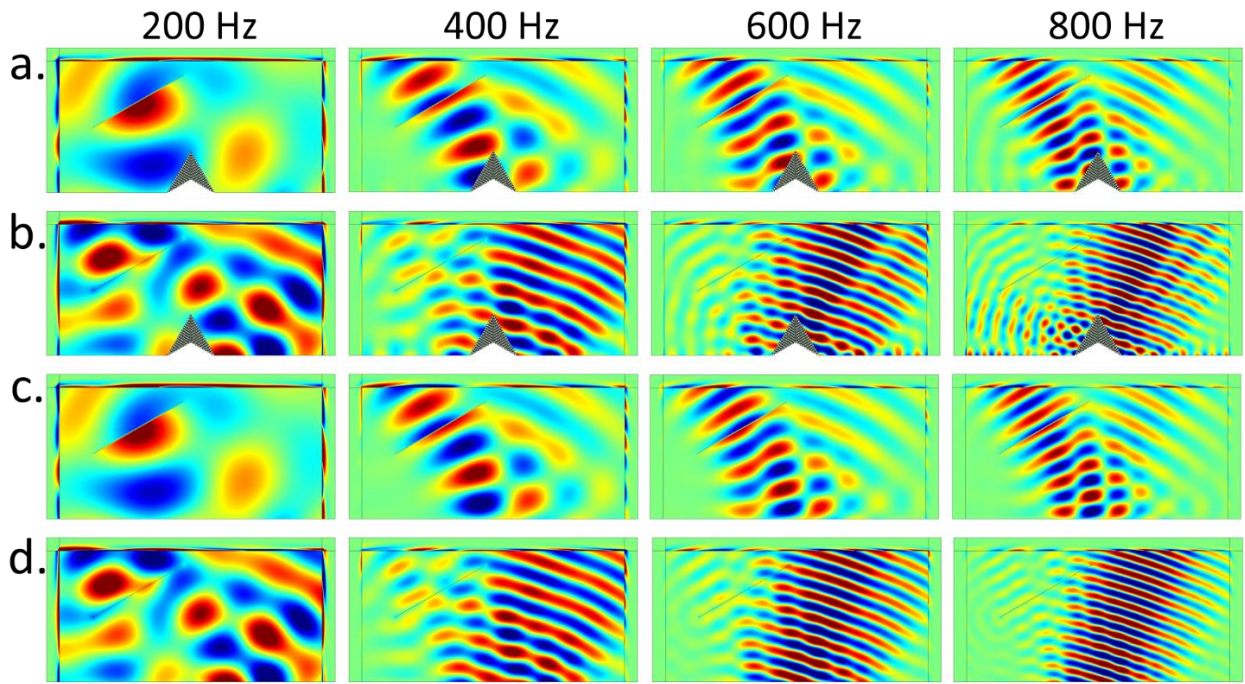

Figure 6. Numerical simulations of the carpet cloak with 30° incidence at different frequencies. Figures in the first to fourth columns show the results at 200, 400, 600, and 800 Hz, respectively. (a,b) Divergence (a) and curl (b) of the displacement field with a pressure incidence, where the cloak is embedded in the background medium; (c,d) Divergence (c) and curl (d) of the displacement field with a pressure incidence, where the background medium is intact.

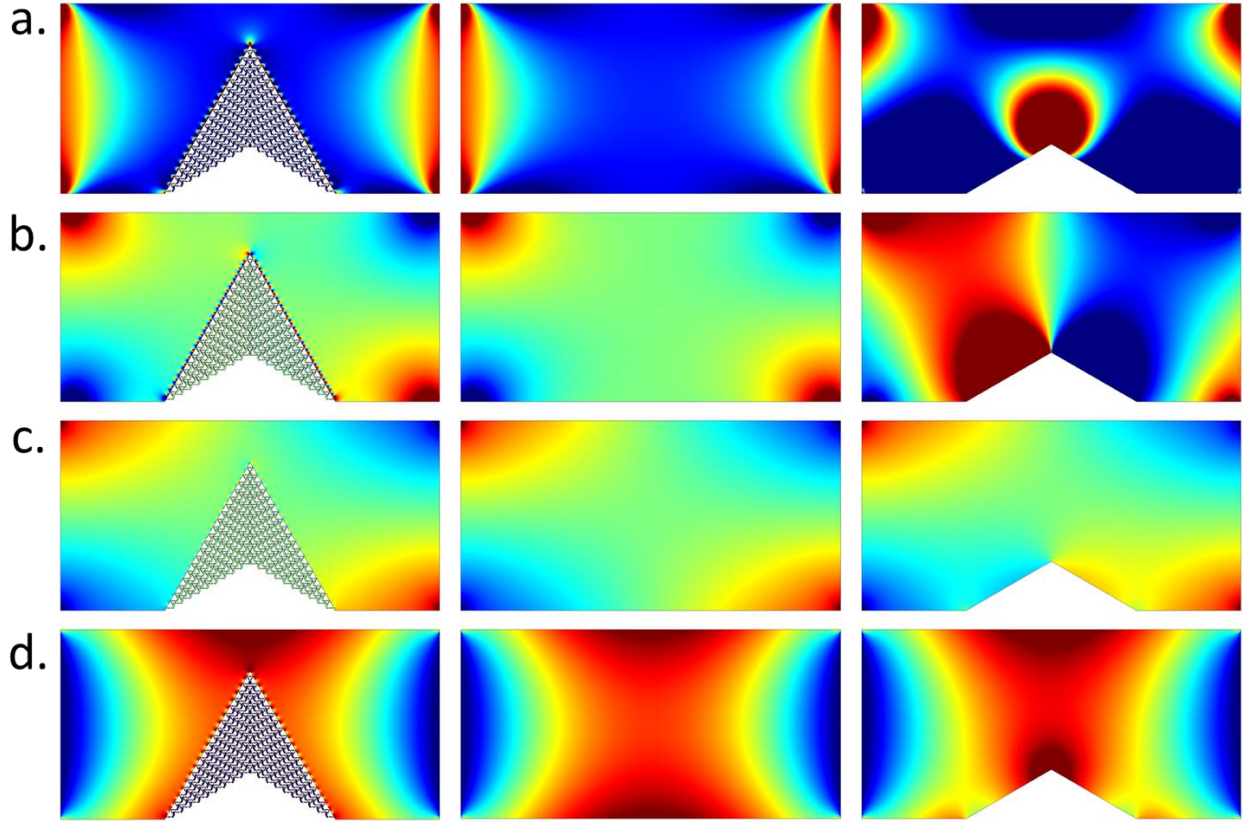

Figure 7. Numerical simulations of the carpet cloak for static tests. Figures in the first column show the results when the continuum lattice is embedded in the background medium for cloaking. Figures in the second column show the results when the background medium is intact. Figures in the third column show the results when the void is non-coated. (a) Divergence of the displacement field with elongation along the horizontal direction; (b) Curl of the displacement field with elongation along the horizontal direction; (c) Divergence of the displacement field with a shear along the horizontal direction; (d) Curl of the displacement field with a shear along the horizontal direction.